**The British Astronomical Association and the Great War of 1914-1918**

**Jeremy Shears**

**Summary**

Marking the centenary of the outbreak of the First World War, this paper considers the effect of the war on the BAA and pays tribute to some of its members who were involved in the conflict.

**Introduction**

The years leading up to the First World War are sometimes depicted as a romantic golden age of long summer afternoons and garden parties, basking in a sun that never sets on the British Empire. Whilst this might be a nostalgic misperception, the majority of the population of Britain was taken by surprise by the events which rapidly unfolded during the summer of 1914 (1). The history books tell us that the inevitable descent of the European powers towards war started with the assassination of Archduke Franz Ferdinand of Austria (1863-1914) in Sarajevo at the end of June. The trigger for Britain's entry into the war was Germany's invasion of Belgium on 3 August. Britain had guaranteed Belgium's neutrality and it therefore issued an ultimatum to the German government that its troops must leave Belgium by 11 pm on 4 August. The ultimatum was ignored and thus by the end of 4 August Britain was at war with Germany. A global calamity of unprecedented proportions soon unfolded which was to last more than four years, during which all aspects of British society were affected in one way or another.

The BAA had celebrated its 21st birthday in 1911 and by that time it had already established an international reputation for the quality of its observational work. As in the rest of British society, there were no thoughts of war at the BAA in the first part of 1914 and indeed, as we shall see, some members were caught by surprise at its outbreak whilst participating in expeditions to observe the total solar eclipse that summer. Contemporary accounts of the affairs of the Association during the Great War in the *Journal* and in the *Memoir* on the BAA's First Fifty Years (2), written in the 1940's, naturally focus on the astronomical progress of the Association and its business affairs. Few details were provided of the impact of the war on individual members, although announcements began to appear in the Journal about officers of the Association leaving for military service and these soon were soon followed by obituaries of members killed in action. This approach was perhaps typical of the era, when the spirit of the age was to serve King and Country, at whatever personal cost, rather than dwelling on its human consequences, which in many cases were too horrendous to consider openly. Moreover, once the war was over, people naturally wanted to forget about it, rather than reliving its horrors (3).

I have written this paper to mark the centenary of the start of the First World War. I mainly focus on the role played by individual members and present some of the personal stories behind their wartime activities, some deeply moving, some tragic and some uplifting. My intention is to pay tribute to the members who strove to protect our freedoms and to those who made the supreme sacrifice. It does not intend to be an exhaustive survey; given a membership in excess of 900 during much of the war, there were many who served their country, beyond the public gaze, and there were doubtless many acts of heroism that have gone unrecorded.

## Solar eclipse expedition of August 1914

Amongst those taken by surprise by the outbreak of hostilities were members of the British Government Expedition to observe the total eclipse of the sun on 21 August 1914, organised jointly between the RAS and the Royal Society. The party, which included BAA Member P.H. Hepburn as a volunteer assistant, sailed from London to St. Petersburg on 17 July (4). They then travelled overland through Russia, learning of the declaration of war by Austria on Serbia *en route*, arriving at Minsk on 30 July, the day the Russian army was mobilised. They feared that in the ensuing disruption their instruments, which travelled separately, would not arrive. However, on 2 August the equipment was delivered safely to the villa that the Expedition members were occupying a few miles outside the city. They avoided Minsk itself as it was one of the main communication centres in the region and it was becoming increasingly frenetic by the day as more and more troops were passing though. Fortunately they observed the eclipse under clear skies – even more fortunate considering most of nearby Minsk itself was clouded out. Hepburn and the party remained a few days to pack their equipment away and then headed back to England, arriving in Newcastle on 7 September.

Another BAA member, Father A.L. Cortie (1859-1925) of Stonyhurst College, had intended to join the Expedition, however he was precluded from entering Russia due to a law banning Jesuit priests (5). Instead, Cortie and several associates, sailed from Hull to Stockholm on 28 July and then made their way to the Swedish town of Härnösand, from where they had a good view of the eclipse (Figure 1 and 2). Returning home was not quite so straightforward and the party needed to take a more tortuous route. Before their ship reached the North Sea, they had to cross two Swedish minefields. Once in the North Sea they were stopped by a British cruiser and warned of German minefields that lay ahead. They therefore changed course and headed for Scotland, escorted by another cruiser, passing down the Scottish and English coasts before finally being convoyed into the mouth of the Tyne with three torpedo-boats.

A further expedition, organised by the Solar Physics Laboratory at Cambridge including H.F. Newall (1857-1944) and F.J. M. Stratton (1881-1960), headed for the Crimea (6). They too decided that it was advisable to maintain a low profile and remain out of town due to the commotion caused by the mobilisation. With the outbreak of war, the Russian authorities ordered members of a German eclipse expedition camped nearby to leave the country, but they were subsequently taken as prisoners of war in Odessa. Stratton was recalled to England early as he needed to join his Officer Training Corps contingent at Cambridge, departing on 11 August, 10 days before the eclipse (7). As it turned out, he didn't miss anything, however, since the remaining observers were clouded out. They all managed to reach England safely, but once again their journey was disrupted (8).

**The General Mobilisation**

Britain's entry into the war occurred during the long summer recess from BAA meetings and by the time of the first meeting of the new session, on 28 October 1914, its first effects had already been felt. The General Mobilisation began in Britain on 3 August, with telegrams being sent to every reservist with instructions to report for duty (9). At the meeting it was announced that two retired Army officers had already been mobilised: Colonel E.E. Markwick, the current BAA President, and Major F. L. Grant, the Association's Senior Secretary. Thus neither was able to attend the meeting, at which members also heard reports about the events surrounding the August solar eclipse (10).

Colonel Ernest Elliott Markwick (1853 – 1925: Figure 3) was a highly experienced army officer and had served in the Army's Ordnance Department, both at home and in the colonies, seeing action in the Zulu War of 1879 and the First Anglo-Boer War of 1880-81(11). After more than 32 years in the army, he retired to Boscombe on the south coast in 1905. He had planned to devote himself to his many interests, notably the observation of the sun and of variable stars. An original member of the BAA, Markwick directed the Variable Star Section between 1899 and 1909 and became President in 1912. In both roles he was highly effective and immensely popular. However, his retirement was cut short by the mobilisation and, being the patriotic gentleman that he was, he was of course keen to do his duty. He was appointed Assistant Director of Ordnance Stores at the Headquarters of Irish Command in Dublin. This was seen as a slightly less demanding role for a man in his sixties than a posting at the Front. He remained in that capacity for the duration of the war and was finally demobilised in 1919. On 1 January 1919 he was honoured as Commander of the Military Division of the Most Excellent Order of the British Empire (CBE) in recognition of "valuable services in connection with the War" (12). During his time in Dublin the responsibilities of his position meant that he was able to make very few astronomical observations and he only attended one BAA meeting in London, in 1917 whilst on leave.

Since the October 1914 meeting was the Association's AGM, Markwick had been due to give his second Presidential address and to hand over his office to his successor, Rev. T.E.R. Phillips. In his absence, the Vice President, F.W. Dyson (1868-1939), who was also the Astronomer Royal, read a letter from Markwick which began: (13)

"Ladies and Gentlemen, owing to my having been called up at the recent mobilisation to enter the duties of a post on the Staff of a large Command, I have been able to devote little, or rather no, time recently to astronomical matters. Consequently I am not in a position to make such remarks as I should have liked on the history of the Association during the past Session".

Dyson went on to read Markwick's address on *Stellar Variation* which he had "prepared some months ago". In summarising, Dyson commented that he was "very glad to be of the slightest service to one who was serving his King and country in that anxious time".

The other absentee at the October 1914 meeting, Major Frank L. Grant, had also led a distinguished military career in the Royal Engineers. His final position was as commander of a contingent of the Officer Training Corps in London from which he retired on 20 December 1913 (14). In the 1890s he had served as the Vice President of the BAA's West of Scotland Branch, which met in Glasgow. In his capacity as BAA Secretary he read many papers on behalf of members unable to attend the London meetings. He also received papers submitted to the Association and in his absence his duties were taken on by E.W. Maunder (1851-1928).

Grant was deployed to the Western Front in 1915 and later that year news emerged that he had been shot through both legs in heavy fighting around Hill 70 near Loos, France. He was invalided back to Glasgow where he was looked after at the Yorkhill Military Hospital. Maunder read a letter from Grant at the December 1915 BAA meeting: (15)

"Dear Mr. Maunder – Many thanks for your very kind note of the 5th [December]. I am keeping very well, and have been out on several occasions recently. In a week or so I am leaving here for a hydro at the seaside, and expect to be able to return to duty at the end of next month. I am glad to see that the Association has been able to carry on in spite of the distractions of war".

The President, Rev. T.E.R. Phillips, told the assembled members that he was sure that they "would be very gratified to learn that their secretary had been serving King and country in this way, and they would wish to send their congratulations to him on his escape and their best wishes for a speedy recovery" (15).

Following a period of convalescence, Grant did go on make a recovery and returned to duties. His original intention in handing his role of Secretary to Maunder was that it would only be a temporary measure, as "when he was called away to the war he did not expect he would be very long absent, but now he thought that even after the end of the war, he might still be required on service for a considerable time" (16). Thus he tendered his resignation in June 1916 and was succeeded by Walter Heath.

In the heady days of August 1914 the popular view in the country was that the war would be over by Christmas. Thus Grant's expectations of a swift return to normal life were typical for the time. However, a different reality emerged as the opposing armies settled into a war of attrition across the trenches of the Western Front, although it was always hoped that the next "big push" would prove decisive and lead to victory. Reflecting a spirit of continued optimism about an early end to the war, at the January 1915 meeting G.F. Chambers (1841-1915) spoke about the circumstances of the forthcoming total solar eclipse of February 1916. He reviewed the various possible observing locations along the track, including the Azores, South America and the West Indies, favouring the latter as offering relatively easy access from Britain and reasonable weather prospects. When the question of the war was raised he commented: (17)

"Of course as regards such matters it was unsafe to prophesy, but his own strong conviction was that the war would be over some considerable time before the month of February 1916. He could not help thinking, putting two and two together, that Germany was rapidly being played out. It was a significant fact, which is not perhaps known to many of those present, that our own Government in the contracts they were making for the supply of materials for war purposes had limited the contracts at present to the month of June, and when one considered that Germany was requisitioning door handles and teapots to obtain copper and was getting very near starvation as regarded food supplies he did not think it required a very sanguine man to imagine that there was a fair prospect of seeing the eclipse of February 1916, with the world more or less at peace".

Quite what privileged information Chambers thought he had access to was not divulged, but of course no BAA solar expedition took place in 1916 (18).

**Patriotism and support for Belgian astronomers**

Clearly patriotic feeling was at a high level throughout the land during the first few months of the war and as hostilities commenced everyone was keen to deliver "a bloody nose to the Kaiser". Patriotic fervour led one member, D.W Horner, to suggest that the BAA should become the BEAA: the British *Empire* Astronomical Association although this idea was not pursued further (19). Members were keen to play their part in supporting the war effort. For example, Miss Helen Metcalfe, of Enfield, Ireland, wrote a letter to the Journal appealing for money to buy warm kit for soldiers including the "Sappers and Miners" of the "Indian Contingent" (20), raising £2 5s, including a donation from Horner (21). She made a further appeal for members to support soldiers in the British Expeditionary Force, allowing her to send "a large number of mittens, socks, cuffs, scarfs, cigarettes, and soup squares, as well as a quantity of tobacco" (22) at Christmastime 1914.

There was much public sympathy in Britain towards the Belgian nation as a result of its neutrality having been violated by Germany. As the invaders moved though Belgium, many Belgian refugees fled to Britain, bringing with them stories of atrocities committed by the German army, which further fuelled support for the Belgian people (23). In a show of solidarity, the new BAA President, Rev. T.E.R. Phillips, invited all refugee Belgian astronomers to attend the November 1914 BAA meeting (24). At the meeting, Félix de Roy (1883-1942; Figure 4), a journalist from Antwerp, described how he had endured the siege of Antwerp and escaped from the city on the morning 8 October just as the main German bombardment commenced, eventually making his way to London. Whilst he had not been a combatant in Belgium he had volunteered as a Red Cross stretcher-bearer during the German attack on Mechelen, south of Antwerp. At one stage he had been captured by German soldiers and held for a few hours before escaping in the mêlée. At the end of his talk, de Roy extended his thanks to "the members of the British Astronomical Association, in the name of his brother astronomers, for the kind reception given to them by the President, and also to acknowledge the very warm and cordial hospitality given to the poor destitute Belgians by the English people generally" (25).

De Roy spent the rest of the war in England and often attended BAA meetings (26). He was a keen variable star observer and in later years became the Director of the Variable Star Section. He had had to leave his telescopes behind in Belgium, but fortunately he was loaned an instrument by BAA member Mrs Fiametta Wilson (1864-1920) which enabled him to continue his variable star observations. For many years de Roy had been a leading light in the Société d'Astronomie d'Anvers (Antwerp Astronomical Society) and he continued to publish the society's *Gazette Astronomique* in England. To show their support, several BAA members joined the society, including W.F. Denning (1848-1931) (27) and Grace Cook, who would often conclude her letters with the affiliation "*Membre de la Société d'Astronomie d'Anvers*" (28).

The speaker before de Roy at the November 1914 BAA meeting was the double star observer Robert Jonckheere (1889-1974) of the Hem Observatory, near Lille, in France (29). He also described his flight from France in the face of the advancing enemy. He put his wife and children on the last boat train leaving for the channel ports on 3 October as the Germans closed in. Jonckheere remained in Lille for a further six days until the authorities ordered all men to leave the town. He then had to make his way on foot for three days and nights to Boulogne, following a tortuous route to avoid the advancing German lines before catching a ferry to England (25). As a result of his reputation as a gifted observer, Jonckheere soon found employment at the Royal Observatory, Greenwich, which continued until the end of the war (30).

The occupying German soldiers used the Hem Observatory as a restaurant and casino (Figure 5). Shortly before the German retreat at the end of the war, the observatory was ransacked and much of the equipment was removed. The equatorial was badly damaged. They fired five rounds at the dome, but all missed. It was several years before the observatory was restored.

## Rev. Martin Davidson and Patrick Hepburn

Having already seen Majors Markwick and Grant depart for war service at the time of the General Mobilisation, during 1915 two prominent section Directors were also called away for service at the Front: Rev. Martin Davidson and Patrick Hepburn, Directors of the Meteor and Saturn Sections respectively.

Martin Davidson (1880-1968; Figure 6) (31) was born in Armagh, winning a scholarship to Queen's University Belfast and graduating BA in 1901. He also took a BSc in mathematics at London University in 1909. His love of mathematics led him to study orbital dynamics and thence the orbits of meteors. He joined both the BAA and the RAS in 1911 and the same year he became Director of the Association's Meteor Section. Davidson was ordained in 1904 and for the next decade was a curate in the East End of London. At the outbreak of war he was appointed Chaplain to the Forces, spending much time ministering to troops on the Western Front from 1915. In his absence the Meteor Section was run by two Acting Directors, Fiammetta Wilson and Grace Cook (32). Their enthusiasm did much to keep the Section alive and they encouraged members who were not away on military service to contribute observations. They even suggested that those who didn't normally observe meteors might like to take up the challenge, with the aim of maintaining as complete an observational record as possible. Davidson made a rare appearance at the December 1917 BAA meeting whilst on a fortnight's home leave. With considerable understatement he commented during the meeting on his attempts to observe meteors from the Western Front: (33) "There the difficulties in the way of meteor observation were insuperable: there was no adequate means of distinguishing between a fireball and a star shell".

Davidson was released from his military duties in early 1919 and took up again the directorship of the Meteor Section. He went on to be BAA President in 1936-39 and Director of the Comet Section 1939-45. His support for both the BAA and the RAS during the difficult days of Second World War was remarkable, especially considering that he was ministering to a Parish in the target area of London's Docklands. He received a Doctorate from Queen's University Belfast in 1931.

Patrick Henry Hepburn (1873-1929; Figure 7) , whom we met earlier in connexion with the 1914 solar eclipse expedition to Minsk, was a lawyer by profession, with a London University LLB degree (34). He was especially interested in Saturn which he observed regularly at the Observatory of the Hampstead Scientific Society, which he joined in 1910 (35). He was appointed BAA Saturn Section Director in 1912. Hepburn was an extremely active and energetic individual thus it came as no surprise when he volunteered for military duty in the early stages of the war, even though he was beyond military age. He was not satisfied until he received a Commission as an Officer in the Royal Naval Air Service (RNAS). In line with his taste for adventure he became a kite balloonist, which involved ascending in tethered observation balloons that were employed as aerial platforms for intelligence gathering and artillery spotting. He spent some time around the British coast submarine spotting as he could see several miles out to sea from his elevated vantage point. This was a risky business, since not only was one at the mercy of the weather, but also of enemy fire from the ground and the air – and with an explosive bag of hydrogen located only a few feet away from one's head. Hepburn's coolness helped him on at least one occasion. He and a mechanic made an ascent in a balloon at Richmond, Surrey, with the aim of making some adjustments to the rigging. Suddenly the balloon was caught in a squall, breaking its moorings and turning over completely. Fortunately, the occupants managed to hold on, the balloon righted itself and they eventually came down in Suffolk unscathed.

Hepburn's first overseas tour of duty was to German East Africa in 1916 and as a result of his absence he handed the reins of the Saturn Section to W.H. Steavenson for the duration of the war. Hepburn returned home in mid October 1916. His next tour was to the Mediterranean beginning in November 1917. Initially he spent some time at the port of Taranto in Italy, and then it was on to Malta and Gibraltar. On 1 April 1918 the Royal Air Force was formed and Hepburn was appointed to its No.1 Balloon Training Wing with the rank of Major (36). After demobilisation in 1919, he relieved Steavenson as interim Saturn Section Director and returned to his legal practice, which he did not much enjoy, having considered other career options such meteorology. He much preferred astronomy and in the 1920s was granted permission to use the 28-inch (71 cm) refractor at the Royal Observatory, Greenwich. He also served as President of the BAA 1921-23. He always enjoyed outdoor activities and he died on Christmas Day 1929 having gone missing during a solitary hill walk in the Lake District.

**Captain John Aidan Liddell VC**

Members attending the October 1915 BAA meeting were reminded of the recent death of Captain John Aidan Liddell, VC (1888-1915; Figure 8), at the age of 27. They would already have been well aware of the events as Liddell's remarkable story had been followed by the press for some weeks and he had become both a national hero and a household name (37).

Aidan Liddell was educated at Stonyhurst College in Lancashire where his scientific interests (38), especially astronomy, thrived under the leadership of Fr. Walter Sidgreaves (1837-1919) and Fr. A.L. Cortie. Both were prominent BAA members and at the time Cortie was Director of the BAA Solar Section. Liddell enjoyed using the College's observatory and evidently impressed his mentor as Cortie invited him as his chief assistant on an expedition to observe the August 1905 total solar eclipse in Spain. Young Liddell was placed in charge of the 20 ft coronagraph designed to photograph the sun in eclipse (39) and thanks to his care and diligence he obtained some wonderful photographs (Figure 9). The following year he went on a voyage to South Africa with his parents where he visited the Royal Observatory, Cape of Good Hope, meeting Sir David Gill (1843-1914). Following the visit, Gill wrote to Cortie: (40)

"I found him an exceedingly bright, intelligent young fellow….I think the lad might do far worse than take to astronomy as a profession….I do think this lad has both the taste and the brains for it".

However, Liddell was also interested in wildlife and it was zoology that he decided to pursue at Balliol College, Oxford. After graduation he joined the Special Reserve of Officers in 1912 and was granted a Commission in the 3[rd] Battalion of the Argyll & Sutherland Highlanders. Thus he split his time between training at the Battalion's headquarters at Fort George on the Moray Firth and the family seat of Sherfield Manor, near Basingstoke. When the General Mobilisation came through in August 1914 he first reported for duty at Stirling Castle, before being transferred to a holding camp at Woolwich and then finally shipping out to France where he disembarked on 2 September. He was in charge of a machine-gun section and saw much action on the Western Front, seeing his first encounter with the enemy at Le Mesnil in mid-October. He was awarded the Military Cross for the role he played by advancing his machine-guns and holding back the main German assault (41).

After Le Mesnil, Liddell was moved up to Ypres where he was involved in action in late October. Then it was on to Ploegsteert (familiarly known as "Plugstreet") and the region of Armentières, where he spent November, December and early January. This was trench warfare at its worst, with freezing cold nights, days of torrential rain, lice in his highland kilt and rats running in the trenches. A photograph of Liddell in the trenches is shown in Figure 10. Due to a chronic lack of trained machine-gun crew he had little respite and at one point he was in the trenches for 43 consecutive days without a break. The only light spot was on Christmas Day 1914 when there was an unofficial truce for a few hours in his section of the line. Eventually his long awaited leave came through: a single week from 11 to 18 January 1915. He spent some time at Sherfield Manor and he visited London to shop, but it is telling in view of what he had been through that he chose to spend three days salmon fishing on the River Eden in Cumberland in solitude.

Only too soon his leave was up and he found himself back at the Front. However, within a few days he was feeling unwell and was taken to the battalion hospital. His conditioned worsened and he was diagnosed with influenza and was sent home to recuperate. Thus once again in late January 1915 he found himself back at Sherfield Manor to convalesce, with time on his hands to contemplate the future. He had long been fascinated by motor cars and aviation and had qualified as a pilot the previous June. Perhaps given his experiences in the trenches, he decided to apply for a transfer to the Royal Flying Corps (RFC). Thus, on 4 May 1915 he found himself reporting for duty at Shoreham airfield. After training at Shoreham and Dover, he was awarded his RFC wings in July with the rank of Captain. His field posting was to St. Omer in France to fly RE5 reconnaissance biplanes (42). His first reconnaissance flight was on 29 July during which he was attacked by a German plane that seemed to come out of the blue. The Lewis gun on the RE5 jammed, a common occurrence, leaving Liddell to return fire with his Mauser pistol (43) whilst his observer used a rifle. Neither was terribly effective. During the contact his observer received a surface wound to his hand, but they managed to return to base where they found that the aeroplane had about 11 holes in its canvass.

The weather the following day was poor, so the next flight was scheduled for 31 July which took Liddell and his observer, Second Lieutenant Roland Peck, from St. Omer up the coast, across the front line to Ostend. Returning towards Bruges they were attacked by a German biplane which opened fire. Liddell's RE5 was badly damaged, cutting the throttle cable and smashing the control wheel. He was badly wounded in his right leg and thigh and lost consciousness. The RE5 flipped over and was heading to the ground out of control. Having plunged 3000 feet, suddenly Liddell regained consciousness and fought to get the RE5 flying level; he found that he could just about control the rudder using the control cables. He realised he was badly injured and that he had three choices: to crash land immediately in enemy territory and certain capture, to try and cross the border to neutral Holland where he and his observer would be interned for the duration of the war, or to head for home. Keeping a cool head he chose the latter (Figure 11). They crossed the lines at only about 2800 feet under heavy fire before just making it to the Belgian airfield at La Panne, where Liddell managed to land in spite of a damaged undercarriage.

Peck, who was unharmed, jumped out, but Liddell insisted on remaining in the aeroplane until skilled medical aid arrived to assist him after about 30 minutes.

During the wait he fashioned a temporary splint for his leg and attempted to stop the bleeding. Eventually he was very carefully lifted from the aeroplane and taken to hospital. A photographer happened to be present who recorded all the events, including a smiling Liddell smoking a cigarette being carried away on a stretcher (Figure 12) and within a few days the photographs were posted across the pages of the British newspapers (44), causing great excitement and admiration back home.

Initially Liddell responded well to hospital treatment, but then infection and fever set in. In an attempt to stop the infection spreading further, a Belgian surgeon removed his damaged leg on 18 August. When the news reached Sherfield Manor, his mother Emily Liddell left for Belgium to look after her son. When she arrived, she found Liddell in good spirits has he had just received news that he was to be awarded the Victoria Cross for his bravery. Unfortunately, shortly afterwards another infection set in and he began to deteriorate rapidly, sadly passing away on 31 August.

Liddell's body was returned to England where it was buried at Basingstoke. A memorial mass was subsequently held at Stonyhurst where they honoured their former pupil with pride.

## Lieutenant John Earle Maxwell

Prior to presenting his October 1916 Presidential Address on long-period variable stars, Rev. T.E.R. Phillips took time to comment on the state of the Association and the war. He noted (45) that "we are glad to know that so many of our members, including officials, Directors of Observing Sections, and workers, are doing their duty gallantly in the field, or in other ways serving the cause of humanity and freedom. And of those who have laid down their lives in the war, I would say, we are honoured in their deaths, and we are sure they have not made the supreme sacrifice in vain". He went on to say that he had hope in the "brightening assurances of victory".

During the year, at least four Observing Sections – Jupiter, Saturn, Meteor and Photographic - reported a significant reduction in activity due to the war (46). Indeed, of the latter the Director, F.W. Longbottom (1850-1933), said that "work was almost at a standstill". The librarian, A.M. Newbegin, also reported a significant reduction in the number of books and lantern slides borrowed; moreover he had needed to modify the library opening times to accommodate his voluntary work amongst the wounded in London hospitals.

In the April 1917 meeting, the worrying news was announced that member Lt. J.E. Maxwell (Figure 13) had been reported as missing in action since 30 March (47). Against all hopes, his death was confirmed in June (48). He was 24. John Earle Maxwell (1892-1917) was educated at Haileybury and it was there that he became interested in astronomy, using the school's 6-inch (15 cm) and his own 3-inch (7.5 cm) refractors to observe the night sky. He continued his interest after leaving school and was a frequent visitor to the Hampstead Scientific Society's Observatory, where he met Patrick Hepburn. Maxwell's particular interest was Mars and he wrote an article and several letters on it in *Knowledge* magazine. He joined the BAA on 26 November 1913 (49), proposed by Patrick Hepburn and W.H. Steavenson, and he enthusiastically participated in its meetings. He was elected FRAS in January 1915.

On leaving school, Maxwell intended to pursue a legal career. However, after two years of studying law he changed direction and enrolled as a medical student at Guy's Hospital in London, in January 1913. Less than two years into the course a further dilemma presented itself as he was torn between continuing his medical studies, knowing that there was a shortage of doctors, or joining the military as a combatant where he could play a direct and immediate role in progressing the war effort. He elected to pursue the latter and in January 1916 he gained a Commission in the Royal Naval Volunteer Reserve, being attached to the RNAS as an airborne observer. After basic training in England he headed to the Eastern Mediterranean in the winter of 1916. On 30 March 1917 he set out with his pilot on a long-distance reconnaissance flight over Bulgaria, but they failed to return. Initially they were listed as missing, but news eventually came through that the aeroplane had been shot down on the day he went missing and that both he and the pilot had been killed (50).

Another BAA member who was also an aviator was Gerald Merton (1893-1983), although he did not join the Association until after the war. He later went on to become President (1950-52) and Director of the Comet Section. Merton obtained a BA in Natural Sciences in 1914 and at the outbreak of war he volunteered for the Royal Flying Corps (RFC). On discovering that the RFC only accepted qualified pilots he went away and learned to fly. The following January, with his pilot's license in hand, he enlisted. Initially he was a reconnaissance pilot in France, but the bulk of his service was in Mesopotamia. He was awarded the Military Cross. He was wounded in 1917 and after hospitalisation in India he was mainly involved in photographic reconnaissance. One story which he liked to tell in later years was how had dinner at an astronomy conference on the 1920's with his one-time German prisoner of war Dr Kruse of Bergedorf Observatory at Hamburg and with whom he had shared his birthday parcel on the day Kruse taken prisoner (51).

### R.F. Roberts and the bombing of London

As the war progressed, the importance of air power began to be realised as a way of attacking the enemy well behind the front lines and Germany had British cities in its sights. The first German air raid took place on 19 January 1915 when two Zeppelin airships bombed the Norfolk towns of Great Yarmouth, Sheringham and King's Lynn, killing four people. Further Zeppelin raids took place during the war and they were complemented by aeroplanes from 1917. The first bombing raid by aeroplanes was on London on 25 May and another on 5 June. This was followed by the first daylight raid on 13 June.

Twenty-two Gotha bombers took off to cross the channel on the morning of 13 June (52). Two were soon forced to turn back with engine problems, whilst another made a diversionary attack on Margate before regrouping with the rest before heading towards London. The first group of bombs was dropped between East Ham and the Royal Albert Docks. The main attack then commenced in which 72 bombs were released within a radius of a mile of Liverpool Street Station between 11.40 and 11.42. There were then raids in other parts of east London, before the raiders headed safely back across the channel. The casualties for London were 162 dead, including 18 children killed by a bomb falling on a primary school in Poplar, and 432 injured making this the worst single air raid of the war. The raid instilled fear and anger up and down the land for the way in which civilians, especially schoolchildren, were targeted.

Following the raid, the body of R.F. Roberts was found inside a doorway at his offices at Fenchurch Street (53). Richard Frind Roberts (1849-1917) ran a firm of chartered accounts the premises of which were almost completely destroyed during the main bombing wave. Several employees were also killed, including his daughter, Lilian Rebecca Roberts (1879-1917) , who was assisting him in clerical duties as so many staff were absent on war duties. Roberts was a keen amateur astronomer, being elected as a member of the BAA in November 1899 and as a Fellow of the RAS in 1906. He owned a 3.7-inch (9.4 cm) refractor which was primarily used for solar observations. It was fitted with a small spectroscope that had been specially designed for him by his close friend, John Evershed (1864-1956). Evershed, who also supported his RAS application, was a leading professional astronomer of the time. A solar expert, he used spectroscopy to discover the 'Evershed effect', the motion of gas in sunspots and was at one time the Director of the BAA Solar Spectroscopic Section. Roberts joined two eclipse expeditions: one to Algiers in 1900 and another to Majorca in 1905. His observations of the 1900 eclipse, which he observed from the roof of the Hotel Continental (Figure 14) along with his son, were reported in the BAA *Memoir* of the event (54).

Sir Frank Dyson announced the loss of Roberts and his daughter before giving his Presidential Address at the October 1917 AGM. In informing members he said: (55) "We deplore their deaths, and would extend our sympathy to his relatives. Their deaths and those of other civilians will not have been in vain if they confirm the country's resolve that the war must be pursued till these and other German crimes are made impossible".

The Roberts family lived in Warlingham, Surrey. A stained glass window was placed in the east wall of All Saints Church, Warlingham, to commemorate Roberts and his daughter. (56)

Air raids continued until May 1918. An aurora on the night of 7 March 1918 coincided with a heavy air raid on London and there was some confusion between the aurora itself and the orange glow due to the burning buildings. It was reported that during the night one "local fire brigade had turned out to extinguish a rosy-coloured aurora" (57).

## Captain A.F.B. Carpenter VC and the Zeebrugge raid

A prominent BAA member in the early 1900s was Captain Alfred Carpenter DSO (1847-1925), a retired Royal Navy officer (58). He joined the Association in 1900 and presented several papers at its meetings. In 1916 he described a navigation diagram that his son, Commander Alfred Francis Blakeney Carpenter (1881-1955; Figure 15), had developed to aid accurate navigation by the stars (59).

The BAA meeting on 1 May 1918 was unusual in that it was given over to a series of short talks rather than the normal papers. Carpenter senior gave a short address on the recent exploits of his son, who was not a member of the Association, as Captain of HMS *Vindictive* to a rapt audience (60). *Vindictive* was a cruiser and was obsolescent by the beginning of the war, but in early 1918 she was fitted out for her final act: the Zeebrugge Raid of 23 April (Figure 16). Zeebrugge had become a key port for German U-boats which were wreaking havoc on British merchant shipping and the Royal Navy intended to neutralise it by blocking the entrance to the navigation canal by sinking ships there. *Vindictive* was prepared by removing all unnecessary superstructure and equipment, including her masts, and fitting her with 7.5-inch Howitzers, flame throwers and mortars. She and the other ships rendezvoused south of Clacton and at the time most of the fleet were unaware of their target. *Vindictive* led with a diversionary attack on the Zeebrugge Mole, guided by the navigation diagram Carpenter had devised, during which she came under heavy fire from German guns. Carpenter "showed most conspicuous bravery, and did much to encourage similar behaviour on the part of the crew, supervising the landing [of 200 troops] from the *Vindictive* on to the mole, and walking around the decks directing operations and encouraging the men in the most dangerous and exposed positions" (61).  With great pride such as only a parent can show, Carpenter senior described how "My son's clothing was riddled with bullets and shrapnel – four through his cap, one cut away his goggles from round his neck, and one passed through the gauntlet of his glove. A fragment or splinter temporarily disabled his arm, but on leaving, when the helmsman was bowled over, he was able to take the steering wheel until a new hand could relieve him".

The *Vindictive* and other ships were sunk as intended, although not quite in the correct position, so they only managed to block the navigation channel for a few days (62). Nevertheless, the Zeebrugge Raid was promoted by Allied propaganda as a key British victory, not least because it involved a large number of volunteer British ships. Eight VC's were awarded, including one to Carpenter (63). After the war he took several other senior commands, retiring Vice-Admiral in 1934.

**1918 and the Armistice**

Germany's final push on the Western Front was led by General Erich Ludendorff (1865-1937), commencing on 21 March 1918. The aim was to make a decisive blow before US troops arrived. Little progress was made and heavy German casualties contributed to a further reduction in morale on the home front. Life was also becoming more difficult for civilians in Britain, with shortages of basic foods and materials becoming more evident. A national paper shortage affected the BAA as the Controller of Paper ordered the Association's publisher to reduce by half the amount of paper it used. Thus the *Journal* shrank noticeably in size and the publication of *Memoirs* was suspended (64).

An important astronomical event occurred on 8 June 1918: the appearance of Nova Aquilae, the brightest nova of the twentieth century. It was almost as if this marked the beginning of a gradual a turning of the tide of war in favour of the Allied forces. This accelerated with the Hundred Days Offensive which began on the Western Front on 8 August. By the time of the next BAA meeting on 30 October 1918, the President, Sir Frank Dyson, was able to address the audience in optimistic, yet reflective, terms: (65)

"We meet to-day feeling that the great evil with which the world was threatened is being overcome, and that peace with liberty and security for great and small nations is near at hand. During these years of war the pursuit of our peaceful astronomical studies has served to distract our thoughts for short intervals of time from the anxieties and horrors of the war. But the war has never been long absent from our minds, and to many of our members has brought pain and loss. To them we wish to express our respectful sympathy, coupled with our admiration, for the courage and devotion of the men who have suffered or have fallen".

Twelve days later the Armistice was signed and at 11 am on 11 November 1918 — "the eleventh hour of the eleventh day of the eleventh month" — a ceasefire came into effect. After more than four long years the war was over.

**The return of peace**

Although the armistice was signed in November 1918 it wasn't until the middle of the following year that most of the BAA members who had been away on duties had returned home (66). As part of a programme to maintain morale amongst troops still serving in Belgium and France, several members, including the new President, Harold Thomson (1874-1962; Figure 17), were invited by the War Office to give lectures on astronomy to them. Thomson was able to see for himself the battlefields from Ypres to Mons where so much blood had been shed.

As in the rest of British society, normality gradually returned at the BAA. During the war, more responsibility had naturally fallen on its Officers who remained behind. E.W. Maunder must have been particularly glad to see the return of peace as not only had he taken on the Directorship of the Saturn Section and the role of Association Secretary for a period, he was also busy directing the Solar Section, as well as fulfilling his professional responsibilities at the Royal Observatory, Greenwich. Nevertheless he even found time to write a popular book for servicemen, "*The Stars as guides for night marching in north latitude 50°*" (67). Moreover, Frank Dyson, a busy professional astronomer, made a major contribution to keep the Association's business running as normal whilst serving as President – Council minutes and meeting notes of the period are remarkable in that much of the discussion focussed on business of the Association and astronomy, rather than the conflict going on in the wider world. Another senior figure who played a role in amateur astronomy during the War was Arthur Eddington (1882-1944), Plumian Professor of Astronomy at Cambridge University. Eddington, being a devout Quaker, was a conscientious objector and received much criticism for his stance. More he suggested that British scientists should retain contact with their German colleagues so far as was possible. Other astronomers, notably the astronomer H.H. Turner (1861-1930), Savilian Professor of Astronomy at Oxford University and a member of the BAA, argued that scientific links with the Central Powers should cease. Eddington contributed to morale by giving talks to amateur astronomical societies throughout the War (68). He also served as RAS Secretary during the War and received communications from Willem de Sitter (1872-1934) regarding Einstein's theory of general relativity which was published in 1916. Eddington did much to bring this to public attention.

In his October 1919 Presidential Address, marking the first AGM since the end of hostilities, Harold Thomson spoke of his relief about the war being over (69). He noted that the membership of the Association had only dropped slightly, from 945 on 30 September 1914 to 919 at the end of the war, although of course the work of its observing Sections had declined significantly. He did however note some positive aspects associated with the war measures at home, such as "the total absence of light in the streets of many of our large towns gave some of us beautiful dark skies such as we had never been accustomed to and cannot hope for again". He also suggested that even "the air raids had their bright side for the Astronomer, in that they taught people to raise their eyes to the skies, and perhaps for the first time in their lives to take some intelligent interest in such simple astronomical phenomena as the phases of the Moon. Even if he did mistake them for the lights of airships, it is all to the good that 'the man in the street' at last discovered some of the brighter planets and learnt that the beams of light he occasionally saw in the Northern sky were not necessarily those from searchlights".

Thomson's eyes were firmly set on the future: "Now that the victory for which we waited so long has been achieved, we hope that our Association may resume its progress and expand its scientific work". Gradually, normality did resume and the Association's activities and output picked up. The last remnant of wartime restrictions disappeared on 26 January 1926 when the first post-war Annual Dinner was held.

Sadly, the dream that this would be "the war to end all wars" was not to become reality and history went on, in some ways, to repeat itself a mere 20 years later. Thus many members who had served in the Great War answered the call of their country for a second time during the World War Two (70).

**Acknowledgements**

I am most grateful to David Knight, Archivist at Stonyhurst College, for providing a copy of the illustration of the 1905 solar eclipse from the *Stonyhurst Magazine* used in this paper, and to Doug and Julia Daniels who gave permission to use the photograph of P.H. Hepburn from the Hampstead Scientific Society archives. Marian Haward provided details of the commemoration to the Roberts family at Warlingham Church, Surrey.

This research made extensive use of scanned back numbers of the *Journal*, which exist largely thanks to the herculean scanning efforts of Sheridan Williams; similarly I used the scanned archive of the *English Mechanic*, supplied by Eric Hutton. The NASA/Smithsonian Astrophysics Data System was also accessed in the preparation of this paper.

I am indebted to my referees, Lee Macdonald and Richard Baum, for their wise and constructive comments.

**Address:** "Pemberton", School Lane, Bunbury, Tarporley, Cheshire, CW6 9NR [bunburyobservatory@hotmail.com]

**Notes and references**

1. Increasing militarisation across Europe, especially in Germany, was of concern from the turn of the century. One BAA member, A.N. Brown (1864-1934), was a preparatory school teacher and in 1900 he set up a branch of the Navy League at his school. The aim of the League was to promote awareness amongst the British public, especially young people, of the dependency of the country on the sea and that the only safeguard was to have a powerful navy. Brown's 1904 talk on "Our Fleet Today" was prophetic; he drew attention to the rise of Germany as a naval power, which at the time was becoming a national concern.

He concluded with the hope that "it will be a very long time before our ships have to confront an enemy in battle". For further information about A.N. Brown see: Shears J., JBAA accepted for publication (2012).

2. Kelly H.L., "The BAA - The First Fifty Years", BAA Memoirs, 42, part 1 (1989).

3. The author's great uncle, Pte. Frederick Obed Bush, died in France on 8 October 1918, within weeks of the Armistice. I recall seeing the photograph of "Uncle Fred" in my grandmother's house in Bristol when I was a small boy. All I knew was that he was "a brave soldier who died in the Great War". When I lived in Belgium, I tracked down Fred's grave and visited the Commonwealth War Graves cemetery at Busigny, just across the border in France. One day in the mid 1990's I visited it with my young family. We were the first members of Uncle Fred's family to make the pilgrimage.

4. For Hepburn's account of the eclipse expedition see: Hepburn P.H., JBAA, 25, 28-31 (1914). A further account is in: Spencer Jones H., The Observatory, 37, 379-384 (1914).

5. Cortie described the eclipse adventure in: Cortie A.L., Nature, 94, 202-204 (1914).

6. Newall H.F., The Observatory, 37, 384-387 (1914).

7. Stratton served in the Signals Service of the Royal Engineers throughout the war, achieving the rank of Lieutenant Colonel and winning the coveted Distinguished Service Order and the Legion d'Honneur.

8. Many other countries sent observers to the eclipse including, Italy, Spain, France, USA, as well as Britain and Germany. The members of W.W. Campbell's expedition from the Lick Observatory made it back to England, but they still needed to cross the Atlantic, which was itself problematic. Two observers were unable to cross to Rotterdam to catch their ship to the USA, so they ended up travelling steerage on the first available ship. See: Campbell W.W. & Curtis H.D., Popular Astronomy, 23, 1-11 (1915).

9. The detailed plans for mobilisation were laid out in the War Book, which had been drawn up over the previous three years. Also on 3 August, a Bank Holiday Monday, there was a notice in "The Times" Personal column under the heading "German Mobilisation", instructing all Germans who were eligible for service to return home. Thus the London railway stations were packed with young German men heading for the channel ports.

10. Although not yet a BAA member – he was elected in 1919, and went on to become President in 1930-32 – Major A.E. Levin (1872-1939) was also part of the General Mobilisation on 5 August 1914. He had served in the Second Anglo-Boer War and subsequently became a Territorial Army officer. Upon mobilisation, he served a year at Newhaven, then he went to France in charge of the first Electrical and Mechanical Company to be sent to the Western Front. He was mentioned in dispatch in 1918 April, and later on was sent to Italy, where he remained until he was demobilised in 1919. He was BAA Secretary (1922-29) and President (1930-32). Levin's obituary: JBAA, 50, 78-80 (1939).

11. Markwick's life and career are described in Shears J., JBAA, 122, 335-348 (2012).

29. Jonckheere was born in Roubaix, France, to a Belgian father and a French mother, thus he had dual nationality in childhood. On reaching majority he renounced his French nationality, but in 1947 he regained it.

30. Thorel J.-C., Le Ciel d'une Vie: Robert Jonckheere, SARL JMG Editions, ISBN 2-35185-039-4 (2009).

31. Davidson's BAA obituary can be read in: Porter J.G., JBAA, 79, 244-246 (1969). His RAS obituary is in: Steavenson W.H., QJRAS, 10, 283-284 (1969).

32. W.F. Denning was invited to become Director of the Meteor Section, but he declined. Thus at the Council meeting of 24 Nov 1915 Maunder offered to receive the correspondence of the Section. The following March, Cook and Wilson were appointed as Acting Directors (BAA Council meeting, 29 March 1916).

33. Report of the BAA meeting of 5 December 1917: JBAA, 28, 41 (1917).

34. Hepburn's BAA obituary can be read in: JBAA, 40, 166-167 (1930). His RAS obituary is in: Hollis H.P., MNRAS, 90, 366-369 (1930).

35. For an account of the history of the Hampstead Scientific Society's Observatory and Hepburn's involvement, see Daniels D.G. & Daniels J.V., JBAA, 121, 13-18 (2011).

36. Hepburn was married to the poet Anna Wickham (1884-1947). Her biography contains more information on her husband's wartime escapades, as well as their turbulent marriage. The couple separated in 1928. Vaughan Jones J., "Anna Wickham: A Poet's Daring Life", Madison Books, publ. 2003.

37. An obituary of Liddell can be read in: Cortie A.L., JBAA, 25, 408-409 (1915). His detailed biography was published in a fascinating book: Daybell P., "With a Smile and a Wave - the life of Captain Aidan Liddell VC MC", publ. Pen & Sword Books Ltd. (2005).

38. At school Liddell's nickname was "Oozy"; he liked nothing more than "messing about with chemicals and engines".

39. The '20 ft coronagraph' was a device for photographing the Sun in eclipse and not the type of coronagraph that uses an occulting disc to observe the corona and prominences out of eclipse, which was not invented until 1930.

40. Irwin F., Stonyhurst War Record, Stonyhurst, pg 160 (1927).

41. News of the award of the MC came through the following February, 1915: London Gazette, 18 February 1915. The previous day he had been gazetted as having been mentioned is despatches: London Gazette, 17 February 1915.

42. Liddell was somewhat disappointed when he learned he would be flying RE5's as they were not the most up to date machines.

43. Mauser pistols, obtained from captured German soldiers, were popular accessories for British airmen.

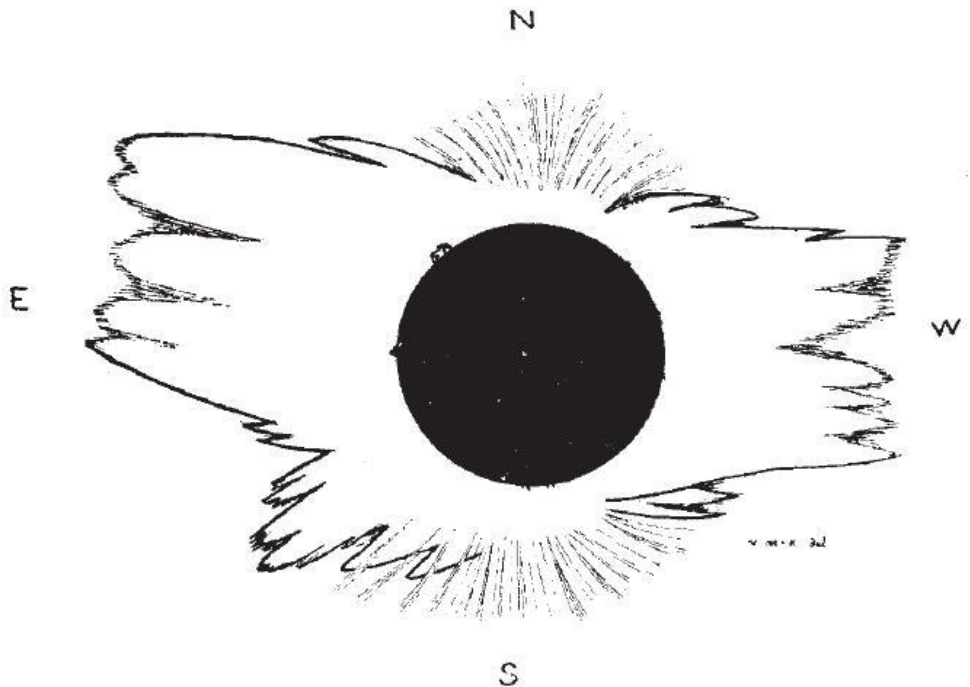

Figure 1: Drawing of the solar eclipse of 21 August 1914 by Father A.L. Cortie

From reference (5)

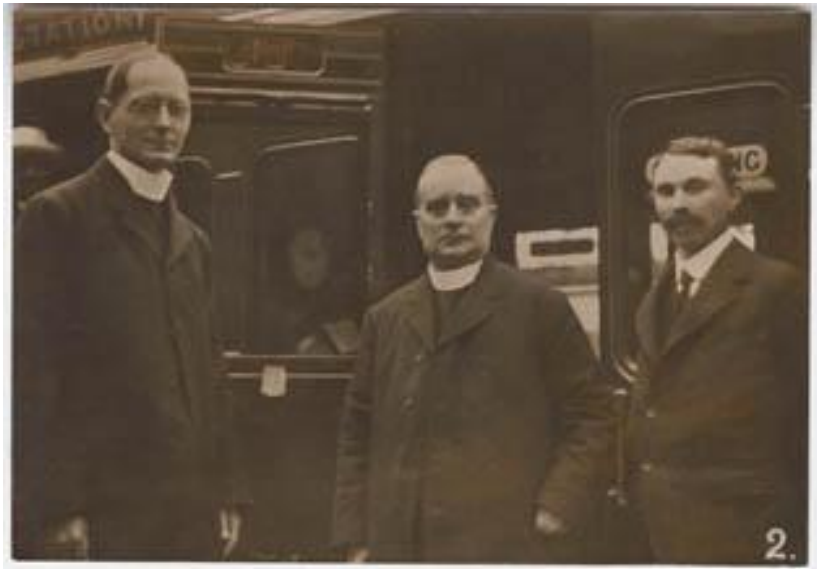

Figure 2: Fr. O'Connor, Fr. Cortie and George James Gibbs *en route* to the 1914 solar eclipse

(University of Central Lancashire)

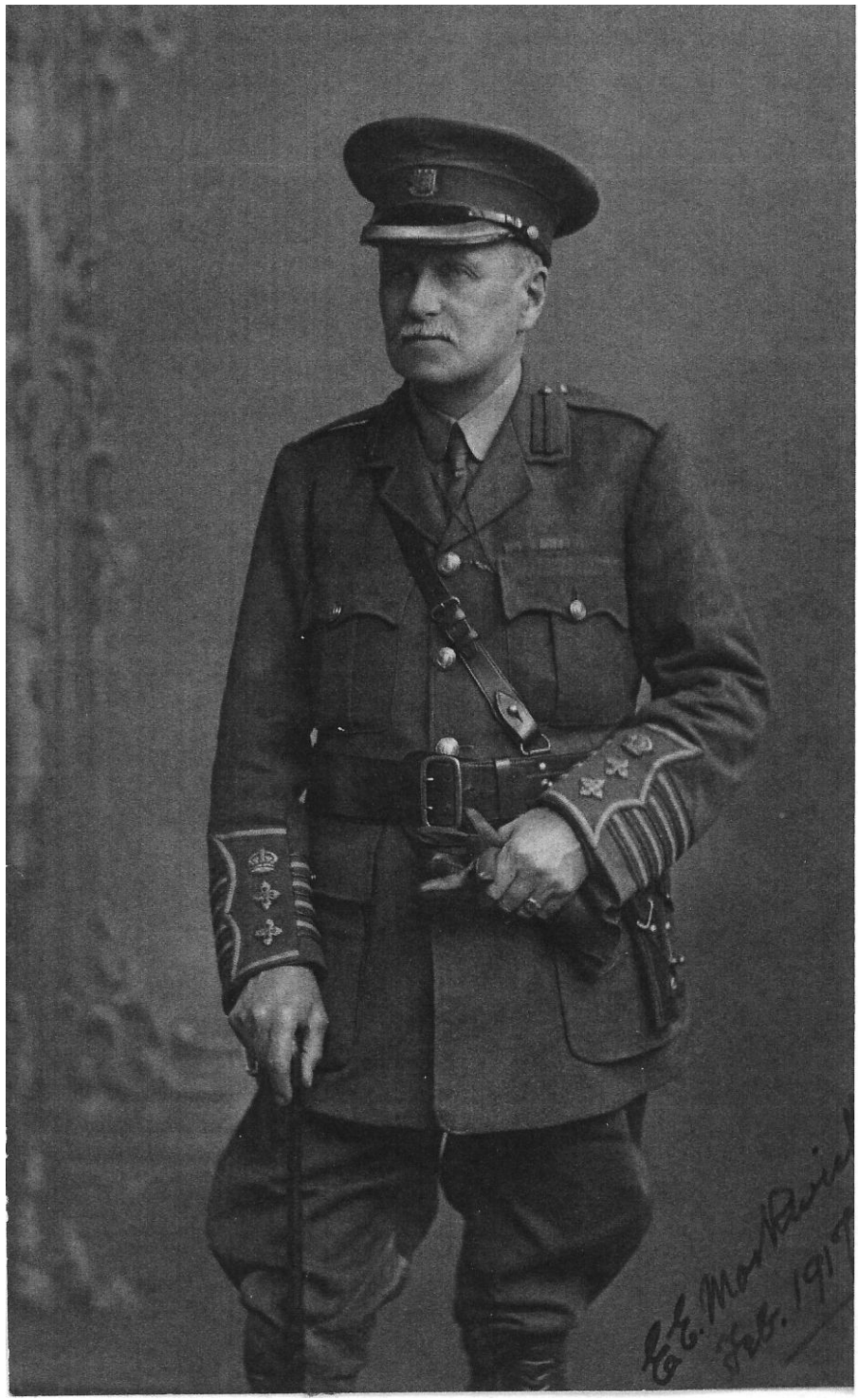

Figure 3: Col. Ernest Elliott Markwick in 1917

(BAA Presidential Portrait)

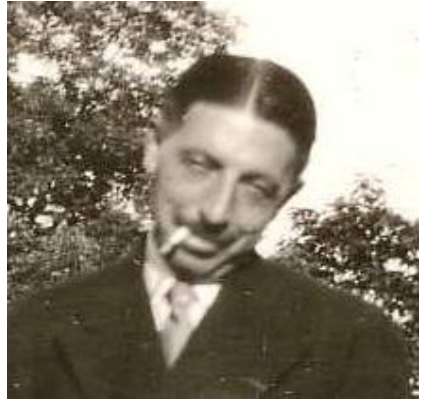

Figure 4: Félix de Roy (71)

(AAVSO archives)

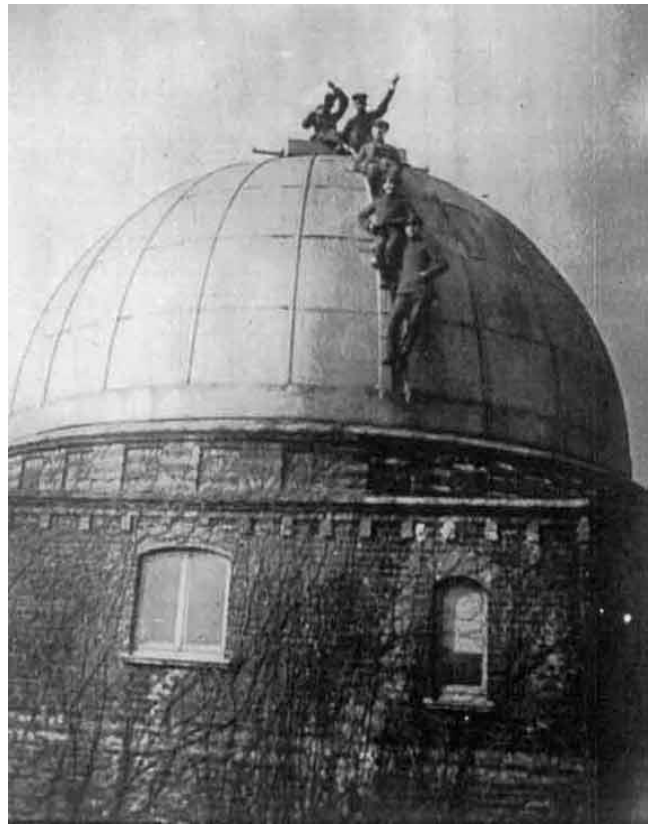

Figure 5: German soldiers pose on the dome of the Hem Observatory, France

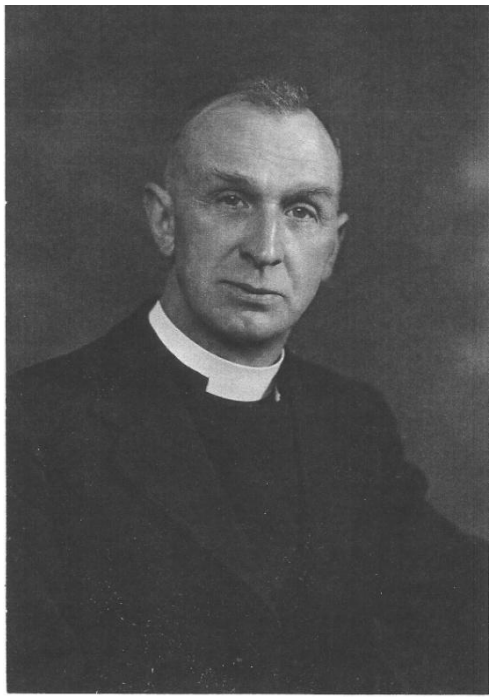

Figure 6: Rev. Martin Davidson

(BAA Presidential Portrait)

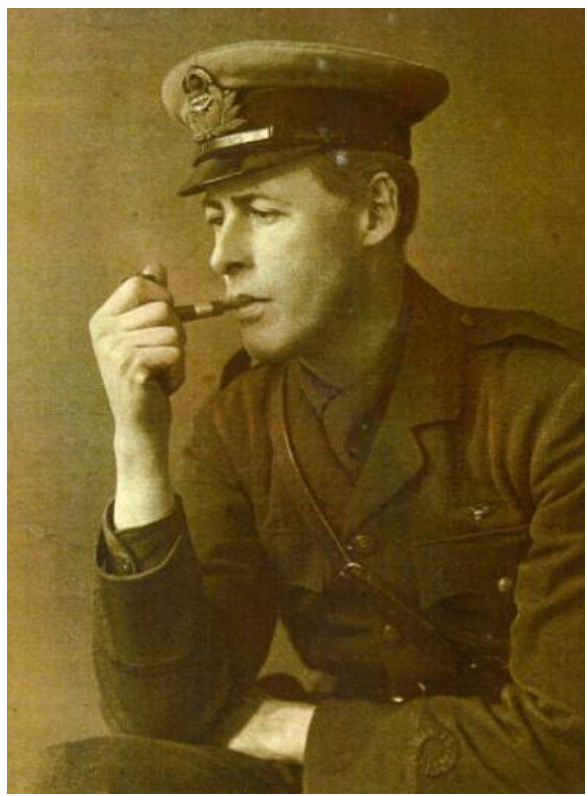

Figure 7: P.H. Hepburn in 1915

(Hampstead Scientific society archives)

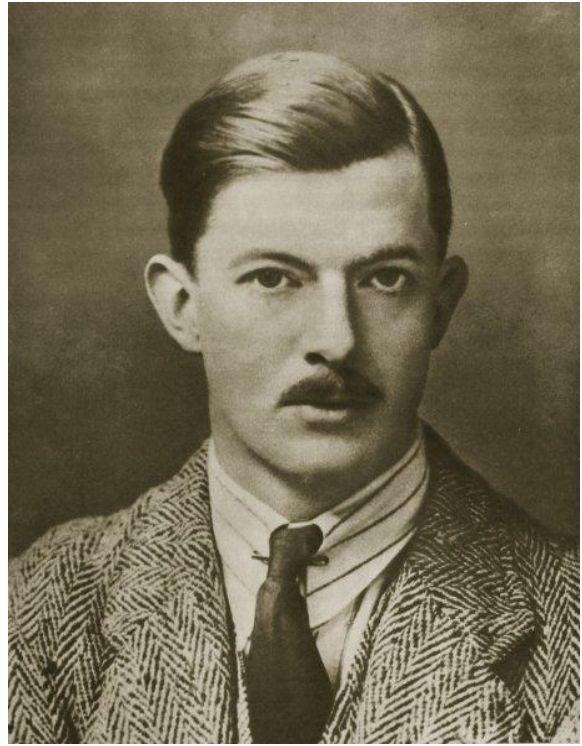

Figure 8: Captain John Aidan Liddell

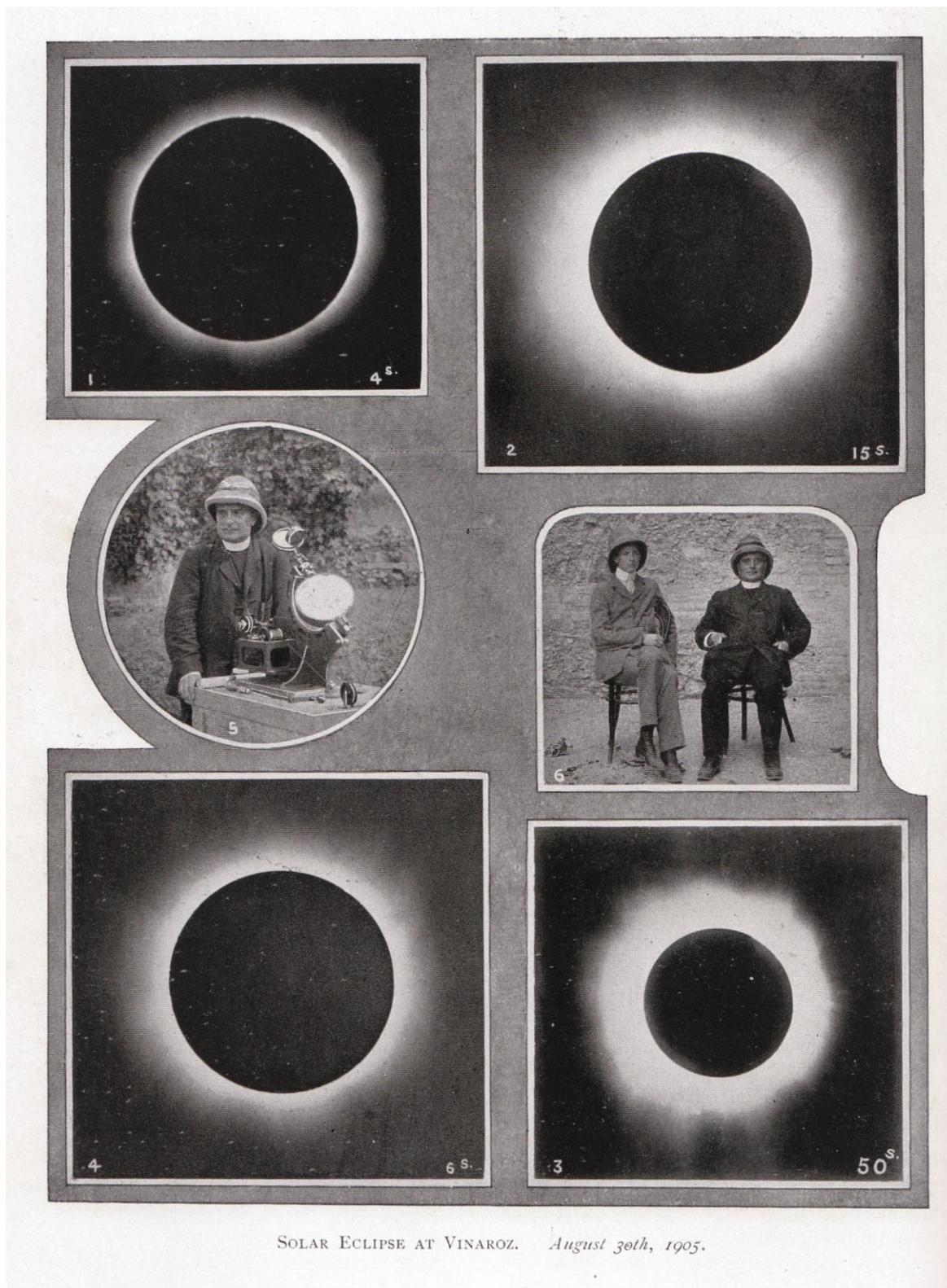

SOLAR ECLIPSE AT VINAROZ. *August 30th, 1905.*

Figure 9: Photographs of the 1905 solar eclipse from the *Stonyhurst Magazine*, showing Fr. A.L. Cortie and the young Aidan Liddell

(Stonyhurst College archives)

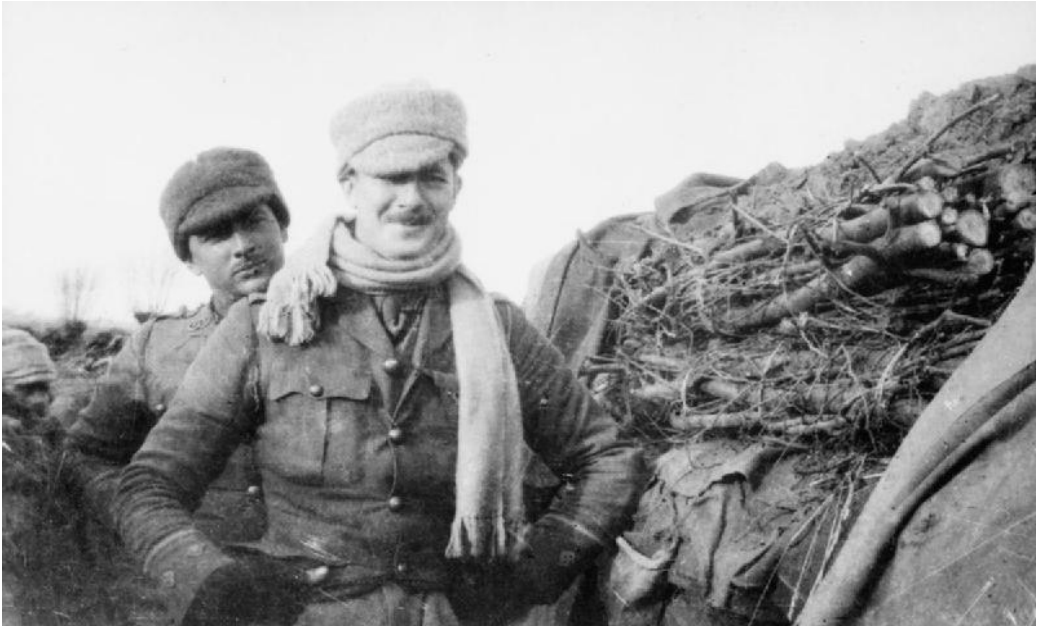

Figure 10: J.A. Liddell during his service with the Argyll & Sutherland Highlanders on the Western Front, 1914 or 15 (Imperial War Museum)

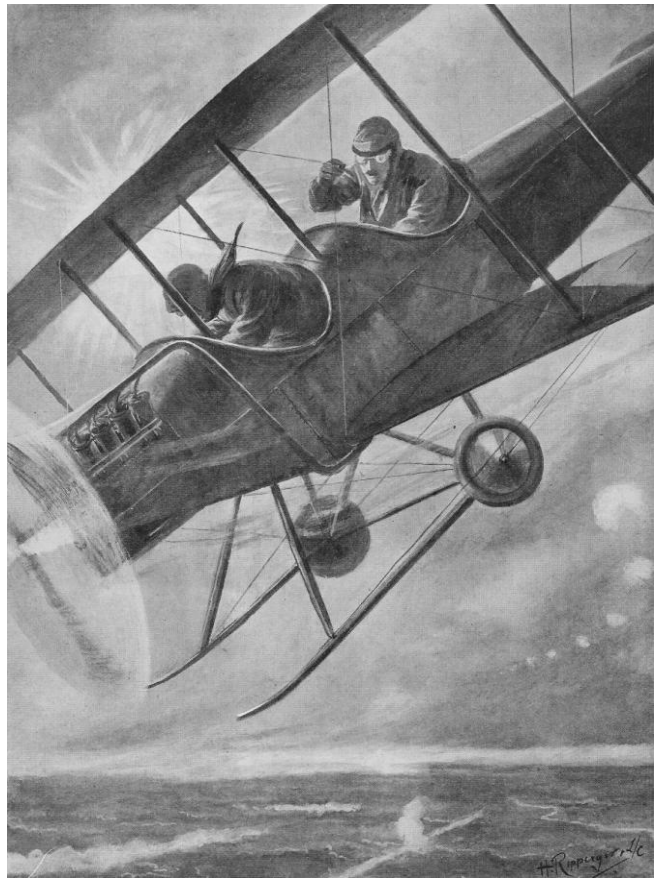

Figure 11: "Captain Liddell Piloting His Aeroplane Down into The British Lines After Being Seriously Wounded". Artist's impression from the popular press (72)

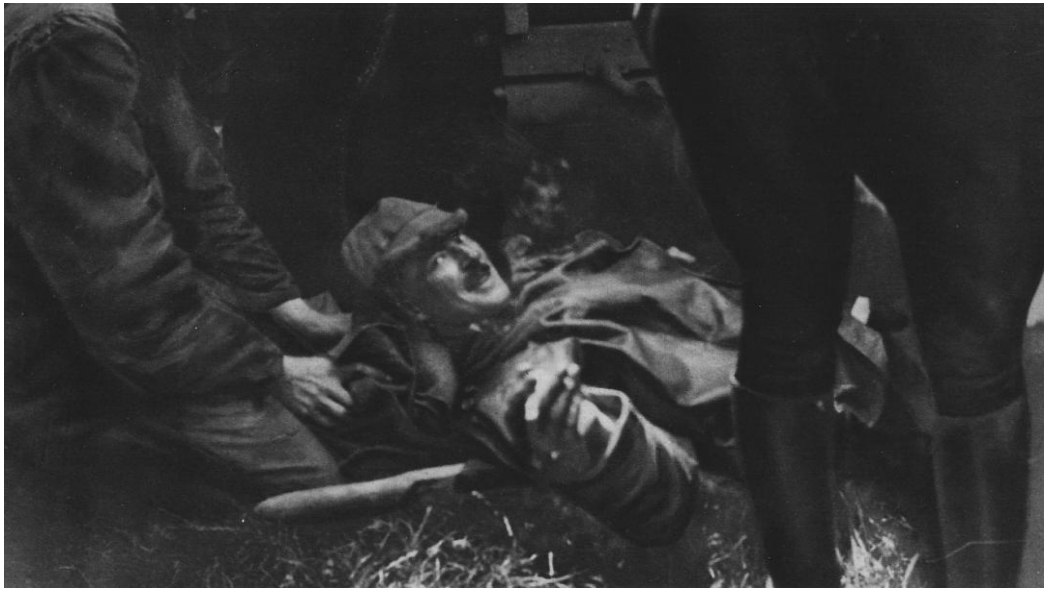

Fig 12: J.A. Liddell just after being taken from his RE5 aircraft (73)

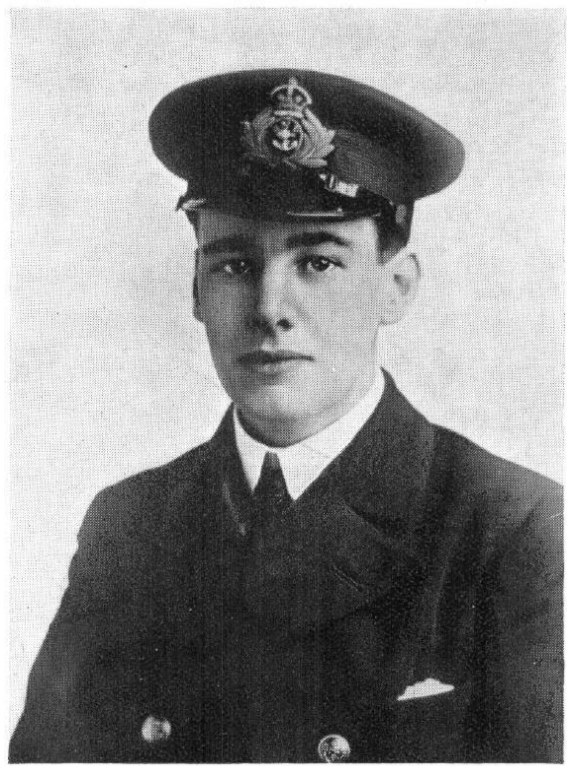

JOHN EARLE MAXWELL.

Figure 13: Lt. J.E. Maxwell

From reference (74)

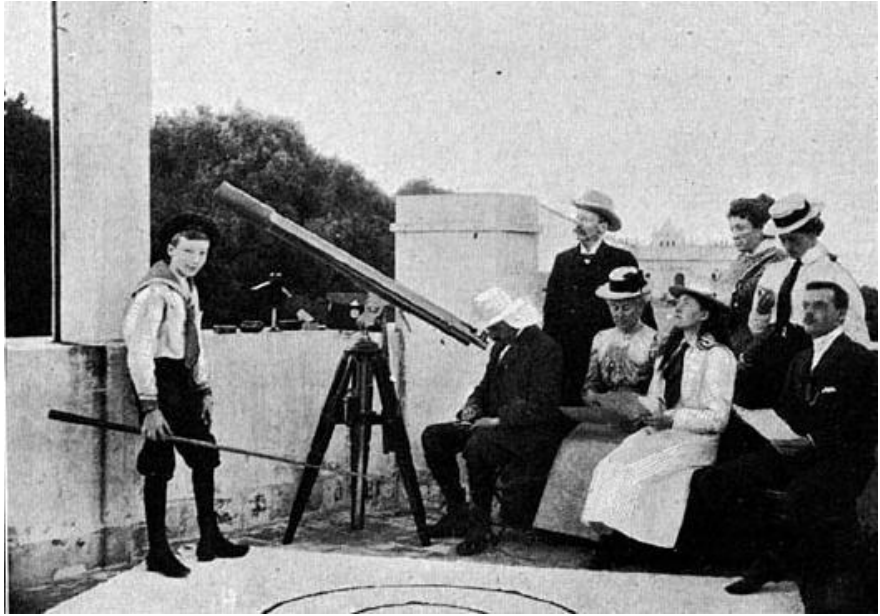

Figure 14: R.F. Roberts viewing through the telescope at the 1900 Algiers eclipse.
His son is seated at the right of the photograph

From reference (54)

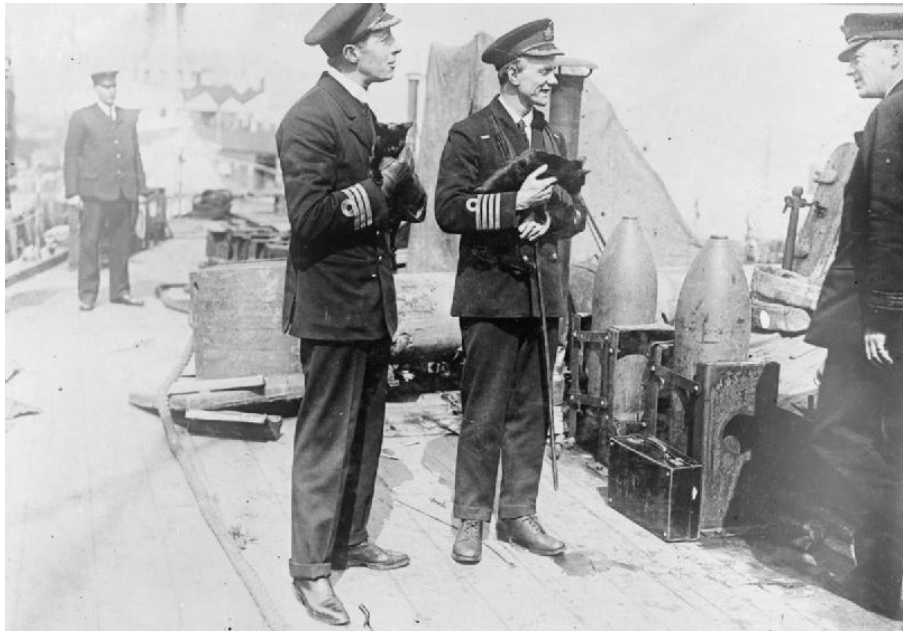

Figure 15: Captain A.F.B. Carpenter VC (centre) on board HMS *Vindictive*, holding
one of the ship's mascots

(Imperial War Museum)

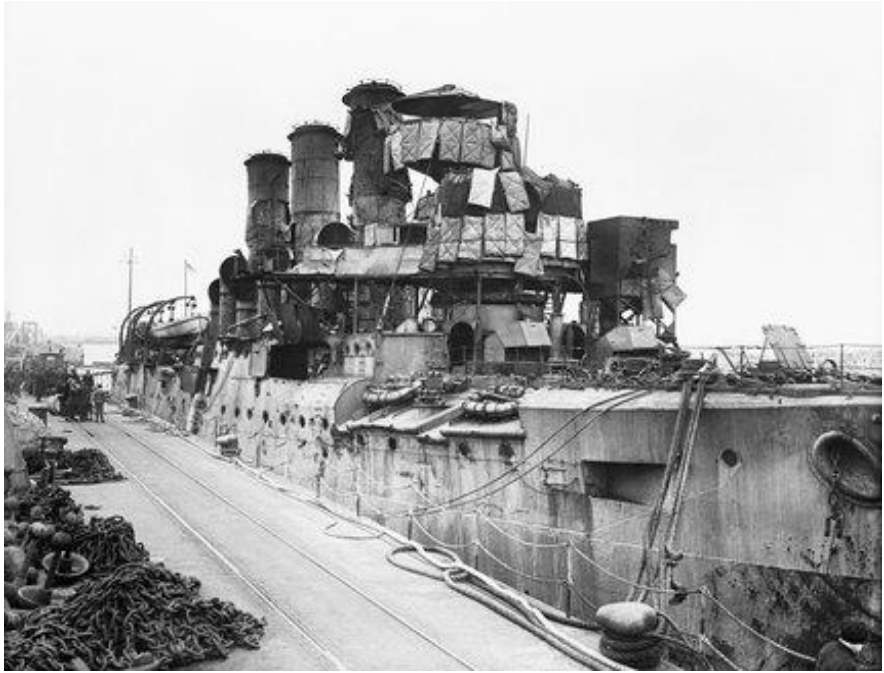

Figure 16: HMS *Vindictive*

(Imperial War Museum)

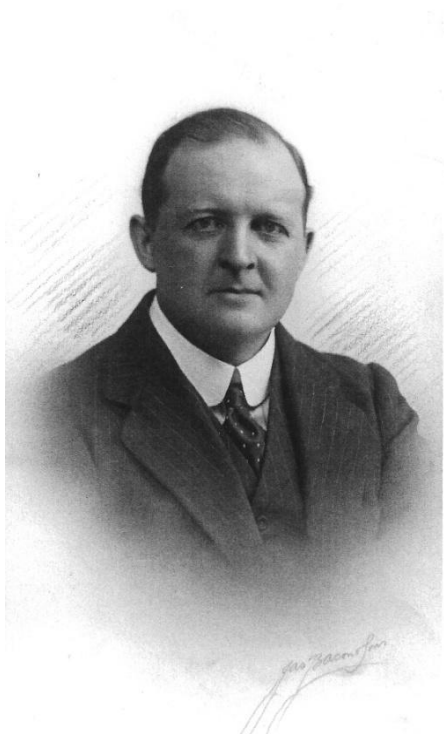

Figure 17: Harold Thomson

(BAA Presidential portrait)